\documentclass{article}

\usepackage{arxiv}

\usepackage[utf8]{inputenc} 
\usepackage[T1]{fontenc}    
\usepackage{hyperref}       
\usepackage{url}            
\usepackage{booktabs}       
\usepackage{amsfonts}       
\usepackage{nicefrac}       
\usepackage{microtype}      
\usepackage{lipsum}		
\usepackage{amssymb,amsmath}
\usepackage{listings}
\usepackage{graphicx}
\usepackage{subfig}
\usepackage{apacite}

\title{Finding the maximum-a-posteriori behaviour of agents in an agent-based model}

\author{
  Daniel Tang\\
  Leeds Institute for Data Analytics\thanks{This project has received funding from the European Research Council (ERC) under the European Union’s Horizon 2020 research and innovation programme (grant agreement No. 757455)}\\
  University of Leeds\\
  Leeds, UK\\
  \texttt{D.Tang@leeds.ac.uk} \\
}

\begin{document}
\maketitle

\begin{abstract}
In this paper we consider the problem of finding the most probable set of events that could have led to a set of partial, noisy observations of some dynamical system. In particular, we consider the case of a dynamical system that is a (possibly stochastic) time-stepping agent-based model with a discrete state space, the (possibly noisy) observations are the number of agents that have some given property and the events we're interested in are the decisions made by the agents (their ``expressed behaviours'') as the model evolves.

We show that this problem can be reduced to an integer linear programming problem which can subsequently be solved numerically using a standard branch-and-cut algorithm. We describe two implementations, an ``offline'' algorithm that finds the maximum-a-posteriori expressed behaviours given a set of observations over a finite time window, and an ``online'' algorithm that incrementally builds a feasible set of behaviours from a stream of observations that may have no natural beginning or end.

We demonstrate both algorithms on a spatial predator-prey model on a 32x32 grid with an initial population of 100 agents.
\end{abstract}

\keywords{Data assimilation, Bayesian inference, Agent based model, Integer linear programming, predator prey model}

\section{Introduction}

The problem we'll consider here is as follows: there is a dynamical system of interest which we believe can be well modelled by an agent based model. We have some prior beliefs about the boundary conditions of the system and have made some observations over some period of time. However, the observations are incomplete (in that they do not specify the full state of the model at any time) and possibly noisy. The problem is to find the most probable sequence of behavioural choices made by the agents (their ``expressed behaviours'') given the observations. The expressed behaviours consist of a feasible set of agent actions or \textit{events} that could have given rise to the observations.

Before presenting an algorithm to solve this problem, we'll first define the problem a bit more precisely.

\subsection{What is an agent-based model?}

We take the essential property of an agent-based model to be that the dynamics of the model can be described entirely in terms of agent behaviours (if the environment contains objects, these should also be modelled as agents). Each agent may have an internal state, and the state of the whole model at any instant can be completely specified by the states of all the agents.

Within this broad definition, there are a number of sub-categorizations
\begin{itemize}

\item Modelled time can be either discrete or continuous. In a discrete-time model, agent behaviour consists of timesteps from time $t$ to $t+1$ and time is an integer. In a continuous-time model agent behaviour consists of discrete events that can happen at any time expressed as a real number. In this paper we will consider discrete-time models. We will also suppose that every agent gets an opportunity to act in each timestep, and that each agent potentially has access to the states of other agents at time $t$ when making the timestep to $t+1$.

\item The internal state of an agent can be discrete, continuous or mixed depending on whether the state can be most naturally expressed as a vector of integers, real numbers or a mixture of both. In this paper we will consider agents with discrete-state. 

\item Agents can be deterministic or stochastic. A deterministic agent will always exhibit the same behaviour if its own internal state and the states of other agents in the model are fixed, whereas a stochastic agent will exhibit different behaviours randomly drawn from a fixed distribution under the same circumstances. In this paper we will consider stochastic agents.

\item If only one agent can occupy a given state at any instant, then the state of the model can be described as a set of agent states and we call it a Fermi-Dirac model, whereas if two or more agents can occupy the same state at the same time then the model state must be described as a multiset\cite{blizard1988multiset} of agent states and we call it a Bose-Einstein model. In this paper we will consider Bose-Einstein models.

\end{itemize}

We define the behaviour of an agent as a computer program which describes a timestep of the agent. In addition to the usual programming constructs, the program may:
\begin{itemize}
\item query the state of the acting agent at time $t$
\item query whether there exist any other agents in the model at time $t$ that satisfy any given predicate on their state
\item create a new agent in any state
\item remove the acting agent from the model
\item modify the state of the acting agent
\item call a function that returns a random number.
\end{itemize}
A timestep of the whole model consists of the execution of the timestepping program for each agent.

Such a program defines a conditional probability distribution over a set of events \cite{staton2016semantics}. Each event represents an agent performing an action that expresses a specific behaviour, which we will write as $(\Phi|\psi,R,\bar{R})$ where $\psi$ is the internal state of the agent before the behaviour is expressed, $R$ is the set of other agents (states) that must be present for the behaviour to be expressed, $\bar{R}$ is the set of agents that must be absent for the behaviour to be expressed and $\Phi$ is the multiset of agents that are the result of the action occurring (including the final state of the acting agent). So, if we say an agent has a behaviour $P(\Phi|\psi,R,\bar{R})$ then when it is in state $\psi$ in an environment containing all agents in $R$ and no agents in $\bar{R}$, it performs an action that results in $\Phi$ with probability $P(\Phi|\psi,R,\bar{R})$.

So, for example, if an agent in state $\psi$ at time $t$ has a behaviour $P(\{\psi'\}|\psi,\emptyset,\emptyset) = 1.0$ it will, with certainty, change its state to $\psi'$ at time $t+1$ . An agent with behaviour $P(\emptyset|\psi,\{\phi\},\emptyset) = 0.5$ in state $\psi$ in the presence of another agent in state $\phi$ will disappear with probability $0.5$.

Without loss of generality, we assume that all agents in the model have the same behaviour. If necessary, multiple types of agent can be modelled by making the agent's type part of its state.

In any timestep of the model a multiset of events will occur, we will call such a multiset a \textit{model event} to distinguish it from the \textit{agent events} that are members of the model event. The consequence of a model event, $E$, is the sum of the consequences of its members. We'll write this as $\Psi(E)$
\begin{equation}
\Psi(E) = \sum_{(\Phi|\psi,R,\bar{R})\in E}\Phi
\label{psie}
\end{equation}
where a sum over multisets is defined in the obvious way. We define an event trajectory, $\mathcal{E}$, as a list of model events $\left<E_1...E_n\right>$, which can be thought of as the assertion that, for each $i$, the events in $E_i$ occurred in timestep from $t=i-1$ to $t=i$. Not all event trajectories are feasible as, for each agent event, the presence and absence conditions of the event must be met.

Given an event trajectory, $\mathcal{E} = \left<E_1...E_n\right>$, and a \textit{seed agent},  $\psi_\emptyset$, let
\[
\Psi_i =
\begin{cases}
 \{\psi_\emptyset\} & \text{if }i=0\\
 \Psi(E_i) & \text{otherwise}\\
\end{cases}
\]
and let $\mathcal{E}$ be \textit{feasible} if and only if $\forall i\in 1...n:$

Each agent must perform exactly one action per timestep
\begin{equation}
\sum_{(\Phi|\psi,R,\bar{R})\in E_i}\{\psi\} = \Psi_{i-1}
\label{agency}
\end{equation}
...each event's presence-requirements must be fulfilled
\[
\bigcup_{(\Phi|\psi,R,\bar{R})\in E_i} R \subseteq \Psi_{i-1}
\]
...and each event's absence-requirements must be fulfilled
\[
\left(\bigcup_{(\Phi|\psi,R,\bar{R})\in E_i} \bar{R}\right) \cap \Psi_{i-1} = \emptyset
\]

The probability of a feasible model event is given by
\[
P(E) = \prod_{(\Phi|\psi,R,\bar{R}) \in E} P(\Phi|\psi,R,\bar{R})
\]
and the probability of an event trajectory is just the product of the probabilities of all the events
\[
P(\left<E_0...E_m\right>) = \prod_{i}P(E_i)
\]

\subsection{What is an observation?}

In this paper we will consider observations that consist of a count of the number of agents for which some predicate is true. The observation can be exact or noisy in that the real count can lie in some range. More formally $\omega = \left<L,U,B\right>$ denotes the observation that the number of agents, $N$, for which predicate $B$ is true is somewhere in the range  $L \le N \le U$. If $E$ is a model event,  then we say that $E$ satisfies the observation, which we write as $E \vdash \omega_i$, if and only if
\[
L \le \left|\left\{\psi:\psi \in \Psi(E) \wedge B(\psi)\right\}\right| \le U
\]
From this, we say that $E$ satisfies a set of observations, $\Omega$, if it satisfies all of its members. We can extend this to model event trajectories $\mathcal{E} = \left<E_1...E_n\right>$ and say that $\mathcal{E}$ satisfies a list of sets of observations (which we'll call an observation of a trajectory), $\mathcal{O} = \left<\Omega_1...\Omega_n\right>$ if for each $i$, $E_i$ satisfies $\Omega_i$.

This form of observation can deal with exact observations ($L=U$), observations where there may be false-positives ($U$ is the observed count and $L=0$), observations where there may be false negatives ($L$ is the observed count and $U=\infty$) or observations that have some uncertainty range ($L = N - \Delta N$ and $U = N + \Delta N$).
 
\subsection{What is a prior belief about boundary conditions?}

The boundary conditions on the system consist of agents entering the system at particular times. At its simplest, it is a set of agents that enter the system at the first timestep and are created by the seed agent, $\psi_\emptyset$, through a behaviour $P(\Phi|\psi_\emptyset,\emptyset,\emptyset)=1.0$, where $\psi_\emptyset \notin \Phi$. More generally, it is a set of probabilities $P(\Gamma,t)$ which give the prior probability that a multiset $\Gamma$ is a subset of the agents that enter the system at time $t$.

In this paper we assume that these prior beliefs can be split into factors that involve single agent states. In this case the prior probabilities can be expressed as
\[
P(\Gamma,t) = \prod_{\psi \in^{=n} \Gamma} P(\psi,n,t)
\]
where $\psi\in^{=n}\Gamma$ is true when $\psi$ appears in $\Gamma$ exactly $n$ times and $P(\psi,n,t)$ is the probability of at least $n$ agents in state $\psi$ being injected at time $t$. 

This can conveniently be represented in terms of the behaviour of a set of hypothetical (though never explicitly represented computationally) \textit{zero energy} agents $\emptyset_\psi$ which, are created by the seed agent $\psi_\emptyset$. The behaviour of a zero energy agent has the form, $(\{\emptyset_\psi, \psi^n\}|\emptyset_\psi,\emptyset,\emptyset)$ which results in $n$ agents in the state $\psi$ being injected into the system, along with the continued existence of the injecting, zero energy agent. If we allow the behaviour of these agents to be time dependent (either implicitly or explicitly via an internal state), then any set of boundary conditions can be represented within the existing formalism.

\section{An algorithm to find the MAP event trajectory}

Bringing all this together, we can now express our problem precisely:

Given an observation of a trajectory $\mathcal{O}$ and an agent behaviour $P(\Phi|\psi,R,\bar{R})$, find
\begin{equation}
\mathcal{E} = \arg\max_{\left<E_0...E_m\right>}\prod_{i}\prod_{\epsilon\in E_i, }P(\epsilon)
\end{equation}
subject to $\mathcal{E}$ being feasible and satisfying $\mathcal{O}$.

To turn this into a tractable algorithm, first represent an event trajectory as a set of non-negative integers $c_{te}$ such that in timestep $t$, the event $e$ occurs exactly $c_{te}$ times.

Now encode each event in an agent's behaviour, $e = (\Phi|\psi,R,\bar{R})$, by letting 
\begin{equation}
\Phi_{e\phi} = n: \phi \in^n \Phi\\
\end{equation}

\begin{equation}
\psi_{e\phi} = 
\begin{cases}
1&\text{if } \phi = \psi\\
0&\text{otherwise}\\
\end{cases}
\end{equation}

\begin{equation}
R_{e\phi} =
\begin{cases}
1&\text{if } \phi \in R\\
0&\text{otherwise}\\
\end{cases}
\end{equation}

\begin{equation}
\bar{R}_{e\phi} =
\begin{cases}
1&\text{if } \phi \in \bar{R}\\
0&\text{otherwise}\\
\end{cases}
\end{equation}

Now encode the observation predicates for each event's consequences
\begin{equation}
B_e = |\{\psi: \psi\in\Phi \wedge B(\psi)\}|
\end{equation}
and let the probability of each event be written
\[
P_e = P(\Phi|\psi,R,\bar{R})
\]
finally, let the consequences of a model event be encoded as
\begin{equation}
\Psi_{t\phi} = \sum_e\Phi_{e\phi}c_{te}
\label{stateIndicator}
\end{equation}

and define a set of boolean variables $0 \le b_{t\phi} \le 1$ which satisfy
\begin{equation}
0 \le Mb_{t\phi} - \Psi_{t\phi}
\label{bGEconstraint}
\end{equation}
\begin{equation}
0 \le \Psi_{t\phi} - b_{t\phi} 
\label{bLEconstraint}
\end{equation}
for some \textit{multiplicity} $M$ which can be any number larger than the maximum number of agents in any one state at any one time. This, for example, could be set to the expected maximum number of agents in the model at any one time.

We can now express the problem as:

Maximise
\begin{equation}
P = \sum_t\sum_e \log(P_e)c_{te}
\end{equation}
subject to the observations being satisfied
\begin{equation}
\forall \Omega_t \in \mathcal{O}, \left<L,U,B\right> \in \Omega_t: L \le \sum_e B_e c_{te} \le U
\label{observation}
\end{equation}
...each agent performing exactly one action
\begin{equation}
\forall\phi: 0 \le \Psi_{(t-1)\phi} -  \sum_e\ \psi_{e\phi} c_{te} \le 0
\label{IPagency}
\end{equation}
...each event's presence requirements being fulfilled
\begin{equation}
\forall\phi: 0 \le Mb_{(t-1)\phi} - \sum_e\ R_{e\phi} c_{te} 
\end{equation}
...each event's absence requirements being fulfilled
\begin{equation}
\forall\phi: \sum_e\ \bar{R}_{e\phi} c_{te} + Mb_{(t-1)\phi} \le M
\label{absenceConstraint}
\end{equation}
This can be seen to be in the form of a pure integer linear programming problem, which in full generality is NP complete, but for which there exists algorithms that, in practice, are able to find solutions to problems of sizes large enough to be useful.

\section{Online assimilation}

The algorithm described above assumes that the observations span a finite time window. If the observations span a very long window or originate from a stream which has no specific start or end, we would usually split the stream up into smaller windows and deal with the observations one window at a time. However, if we commit to the MAP event trajectory of one window, and use the end state of this as the start state of the next window, we are in danger of being led down ``garden paths''; i.e. the observations from one window lead us into believing the most probable explanation, only for later observations to show us that a much less probable and completely incompatible explanation is in-fact the case. So, we can't commit to believing an event trajectory for one window because, in the light of the observations from the next window, the trajectory may turn out to be impossible. If this happens the assimilation must stop because there is no event trajectory that satisfies the new window's observations while being a feasible extension of the previous window's trajectory. To solve this problem, we present a second algorithm for use with longer windows and streaming observations.

We begin by defining the idea of a \textit{partial event trajectory} which is like a normal event trajectory except that instead of each agent having to perform exactly one action per timestep, they can perform at most one action per timestep. So, in place of equation \ref{agency} we have
\begin{equation}
\sum_{(\Phi|\psi,R,\bar{R})\in E_i}\{\psi\} \subseteq \Psi(E_{i-1})
\label{partial}
\end{equation}
Intuitively, a partial event trajectory can be interpreted as saying that at least these events happened in these timesteps, but other additional events may have happened concomitantly.

In an partial trajectory, all observations up to the end of the current window are satisfied but only a subset of the agents present at any particular time will be moved forward in time by the next model event, so agents may be left hanging at a time in the middle of the trajectory. This is good because it reduces the likelihood that we will be led down a garden path by delaying any commitment to the exhibited behaviour of the hanging agents until more observations are available.

The assimilation proceeds by finding a partial trajectory for one window, then taking the next window of observations and finding the MAP partial trajectory given these new observations along with the partial trajectory we've already committed to. In order to correctly move hanging agents forward in time, we need to distinguish between the events that we've already committed to and the events we're optimising over. If we let $\hat{c}_{te}$ be the number of events of type $e$ that we've committed to in timestep $t$, then equation \ref{stateIndicator} becomes
\begin{equation}
\Psi_{t\phi} -  \sum_e\Phi_{e\phi}c_{te} = \sum_e\Phi_{e\phi}\hat{c}_{te}
\label{stateIndicator}
\end{equation}

the agency constraint, equation \ref{IPagency}, becomes
\begin{equation}
\forall\phi:  \sum_e\ \psi_{e\phi}\hat{c}_{te} \le \Psi_{(t-1)\phi}  -  \sum_e\ \psi_{e\phi} c_{te}
\end{equation}
and the absence constraint, equation \ref{absenceConstraint}, becomes
\begin{equation}
\forall\phi: \sum_e\ \bar{R}_{e\phi} c_{te} + Mb_{(t-1)\phi} \le M - \sum_e\ \bar{R}_{e\phi} \hat{c}_{te}
\end{equation}

Since observations are fully satisfied by a partial trajectory, the observation constraints, equation \ref{observation}, now only apply to timesteps for which there are no committed events. The presence requirement constraints can remain unchanged.

Even though partial trajectories reduce the probability of us being led down the garden path, they do not eradicate it completely. So, if we end up with observations that are incompatible with our commitments, we rollback the commitments, one timestep at a time, from the most recent until a solution becomes possible. In the worst case, we'll have to roll-back all the way to $t=0$, but in practice the probability of this will be very small for most models.

One final issue with online assimilation is agents that leave the model. When an agent leaves the model it will not be observed again, but the online assimilation will just leave it in the hanging state on the off-chance that it's still around but has evaded observation. At some point we are likely to want to commit to saying that the agent has left the model. To do this we can estimate the probability of the agent evading observation for so long, and when this estimate falls below a certain level, we commit to its departure.

If required, a partial event trajectory can be extended into a complete trajectory at any point by requiring that equation \ref{partial} is satisfied with equality and solving the resulting set of equations.

\section{Experiments with a spatial predator-prey model}

In order to demonstrate the above algorithms, we simulated a spatial predator-prey model on a 32x32 grid with periodic boundary conditions. Agents can be either predator or prey. At each timestep, agents can stay on the same grid-square, move to an adjacent grid-square (i.e. up, down, left or right) or die. Prey can also give birth to another prey on an adjacent grid-square. Prey cannot move onto a square that contains a predator, but in the presence of a predator on an adjacent square they may be eaten and the predator reproduce into the newly vacant square. The probability of each behaviour is shown in table 1. Perhaps surprisingly, even a model with such a simple set of behaviours exhibits quite complex emergent properties that are difficult to predict.

\begin{table}
\begin{center}
\begin{tabular}{llc}
\hline
Aget type & Description & Probability\\
\hline
Prey & die &        0.03\\
 & reproduce &        0.06\\
 & move (up/down/left/right) / be eaten &        0.728\\
 & stay put / be eaten &  0.182 \\
 &&\\
Predator & die  &      0.05\\
 & move (up/down/left/right)&        0.76\\
 & stay put & 0.19\\
\hline
\end{tabular}
\end{center}
\caption{The rates of each behaviour in the predator-prey model}
\label{rates}
\end{table}

Forward simulations of the ABM were performed in order to create synthetic observation data. The initial state of each simulation consisted of 40 predators and 60 prey placed at random positions on the grid with uniform probability. The boundary conditions for the MAP analyses consisted of a fully observed initial state. Observations were made at each timestep and each agent was observed with a probability $\frac{2}{3}$ in each timestep.

\subsection{Offline MAP experiment}

Using the above boundary conditions and observations and using a window length of seven timesteps, we set up an integer linear programming problem as specified above and used Google's OR-Tools\cite{googleortools} library to find the MAP event trajectory. In total 16 simulations were analysed, the mean time taken per simulation to compute the MAP using a single core of a 1.6GHz Intel i5-8250U was 156s.

\begin{figure}
\begin{center}
\includegraphics[width = 12cm]{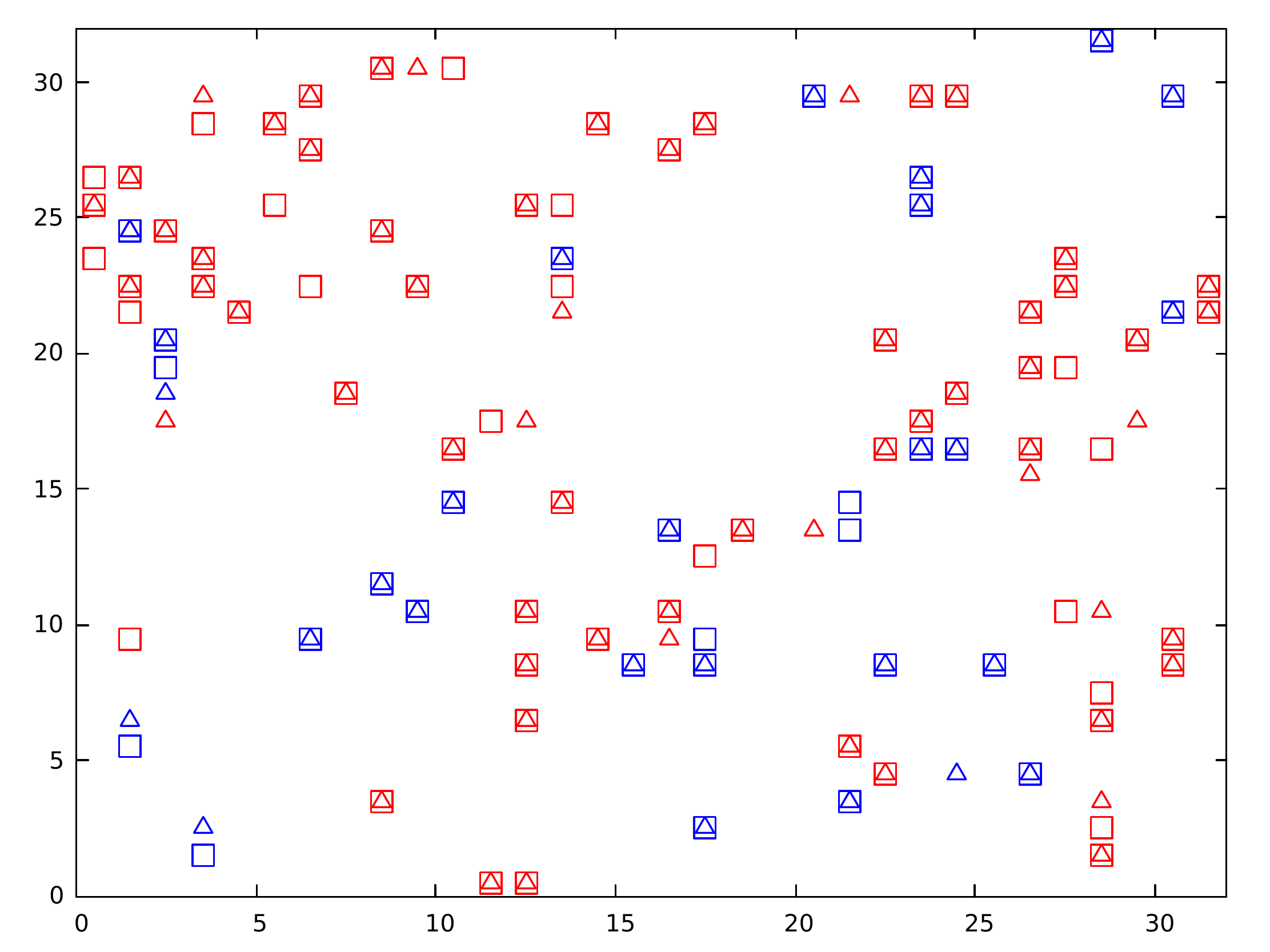}
\end{center}
\caption{Snapshot comparing the real and MAP states at the end of a 7-step assimilation window. Blue symbols indicate the spatial position of predators, red symols indicate prey. Squares show the position of agents in the real state, while triangles show agents in the MAP state.}
\label{snapshot}
\end{figure}

\begin{figure}
\begin{center}
\includegraphics[width = 12cm]{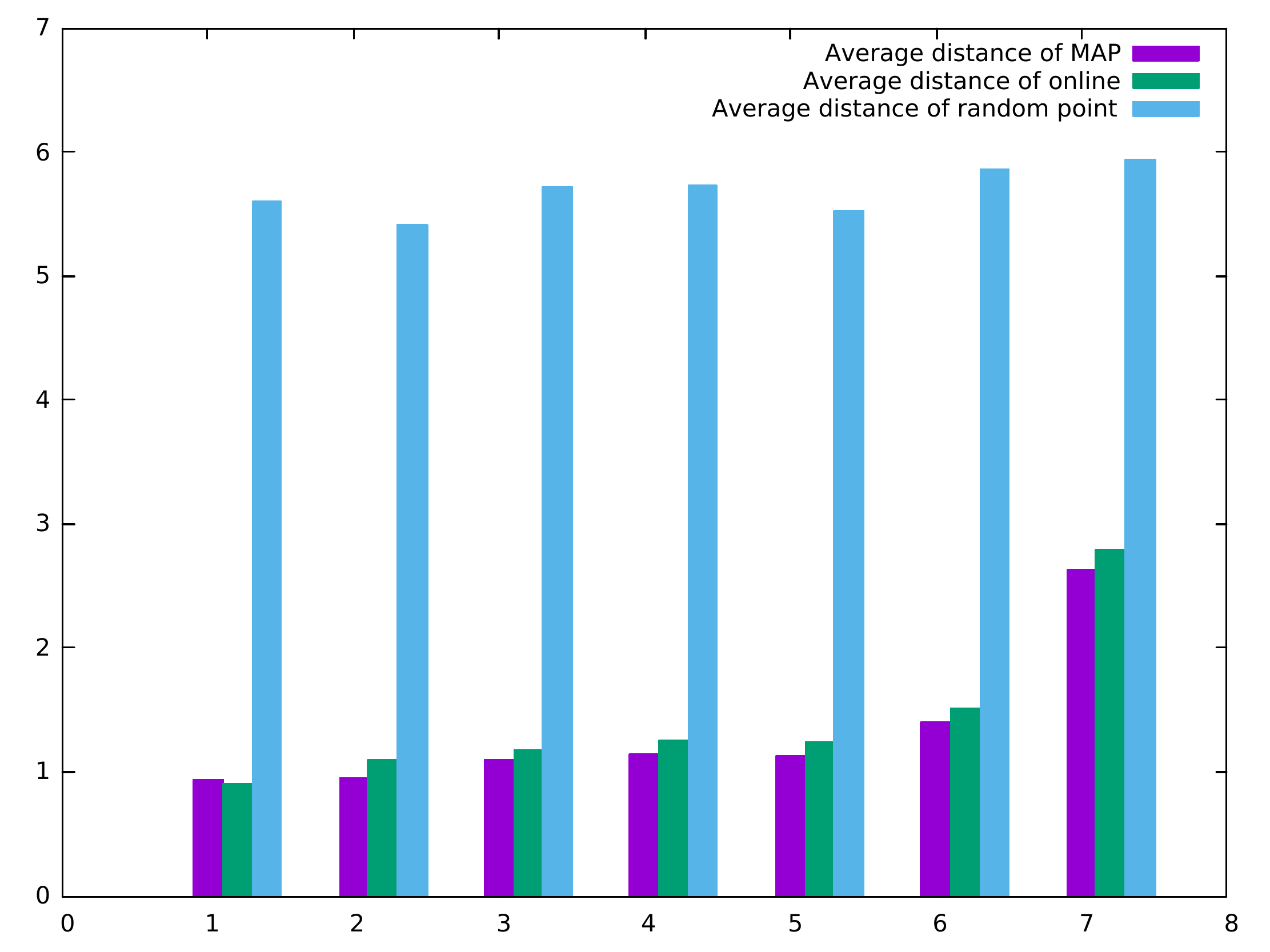}
\end{center}
\caption{The average shortest $L_1$ distance from an unobserved agent in a MAP state to an unobserved agent of the same type in the corresponding real state, plotted against number of timesteps from the start of the window. For comparison, we also show the shortest distance for the trajectories produced by the online algorithm, and the shortest distance for a randomly placed agent.}
\label{distance}
\end{figure}

A comparison between typical real and MAP model states at the end of an assimilation window is shown in figure \ref{snapshot}. As a rough measure of the difference between the MAP and the ``real'' state, Figure \ref{distance} shows the average shortest $L_1$ distance between the unobserved agents in the MAP and the nearest unobserved agent of the same type in the real state for each timestep in the window. For comparison, we also show the average distance for a model that places the agents randomly with uniform probability.

\subsection{Online assimilation experiment}

The online algorithm was applied to the same set of observations and boundary conditions as the offline experiments. A window size of one timestep was used and agents were not removed from the model if unobserved. Once the partial event trajectory for a whole simulation was found, the trajectory was completed by solving one more time for the hanging agents. The mean time per simulation to compute the event trajectory, using the same computer as used for the offline experiments, was 11.4s, representing a speed increase of just under 14 times compared to the offline algorithm.

In order to assess the quality of the online algorithm compared to that of the MAP, we calculated the ensemble average log ratios of the posterior probabilities

\[
\overline{\ln\left(\frac{P(\mathcal{E}_{MAP}|\mathcal{O})}{P(\mathcal{E}_{real}|\mathcal{O})}\right)} = 66.7
\]

\[
\overline{\ln\left(\frac{P(\mathcal{E}_{online}|\mathcal{O})}{P(\mathcal{E}_{real}|\mathcal{O})}\right)} = 44.1
\]

where $\mathcal{O}$ are the observations, $\mathcal{E}_{MAP}$ is the MAP event trajectory, $\mathcal{E}_{online}$ is the trajectory given by the online algorithm and $\mathcal{E}_{real}$ is the trajectory used to create the observations. The MAP log ratio can be thought of as an upper bound on performance while a log ratio of around 0 would be the expected value if we sampled from the posterior. The performance of the online algorithm can be assessed in relation to these two reference points. As can be seen, the online algorithm performs less well than the MAP algorithm, but the online trajectories are, on average, substantially more probable than the real trajectory that was used to create the observations in the first place. 

Figure \ref{distance} shows the average shortest $L_1$ distances between agents in the the online trajectory and those in the real trajectory, alongside those of the MAP trajectory.

\subsection{Code}

The code for all experiments described here is available at \href{https://github.com/danftang/MaxAPosteriori}{https://github.com/danftang/MaxAPosteriori}

\section{Conclusion}

We have demonstrated that the maximum-a-posteriori trajectory of an agent-based, spatial predator-prey model can be found given a set of partial observations. We have also shown how to construct a feasible interpretation of a stream of such observations of arbitrary length.

Since integer linear programming is NP-complete the technique described here will not in full generality scale to very large models but will work best when agents choose from one of a limited number of options at each timestep, when interactions are sparse and when the observations impose strong constraints on the state. However, expressing the problem in this form lays the foundations for further research to apply approximations that may lead to more scalable algorithms.

Finally, it is worth noting that while the MAP event trajectory is the most probable set of events given the observations, it is also usually highly unlikely to be the true cause of the observations, for the simple fact that it is one among a very large number of other trajectories with non-trivial posterior probability. For this reason it is not usually a good idea to infer values of unobserved variables from the MAP. Rather, the MAP should be used as one component of a larger analysis. For example, the MAP trajectory is often a good choice for the start state of a Markov chain Monte Carlo sampling algorithm, which will be described in our next paper.

\bibliographystyle{apacite}
\bibliography{references}

\end{document}